\documentclass[a4paper,11pt]{article}
\pdfoutput=1 

\usepackage{jinstpub} 

\usepackage[utf8x]{inputenc}
\usepackage{upgreek,xspace}
\def\Sigma{\ensuremath\upsigma\xspace}
\usepackage{siunitx}
\usepackage{mhchem}

\title{\boldmath Module concept and thermo-mechanical studies of the silicon-based TT-PET small-animal scanner}

\author[a]{D. Ferrère,}
\author[b]{Y. Bandi,}
\author[a]{F. Cadoux,}
\author[b]{D. Forshaw,}
\author[a]{D. Hayakawa,}
\author[a]{G. Iacobucci,}
\author[a]{S. Michal,}
\author[b]{A. Miucci,}
\author[a,c]{M. Nessi,}
\author[a]{L. Paolozzi,}
\author[a]{E. Ripiccini,}
\author[a]{P. Valerio,}
\author[b]{M. Weber}

\affiliation[a]{DPNC, Département de physique des particules et corpusculaire, \\Geneva}
\affiliation[b]{University of Bern, \\Bern}
\affiliation[c]{CERN,\\Geneva}
\emailAdd{didier.ferrere@unige.ch}

\abstract{The construction of the small-animal TOF PET scanner proposed for the TT-PET project poses several technical challenges, with the need to address tight mechanical, thermal and electrical constraints. The scanner has a length of \SI{5}{\centi\meter}, an inner diameter of \SI{3.9}{\centi\meter} and an outer diameter of \SI{8.4}{\centi\meter}. Each of the 16 towers constituting the scanner is made of a stack of 60 monolithic silicon detectors with excellent time resolution, specifically developed for this project. The interconnections necessary for the read-out and power distribution of the chips are obtained by a stacked-die wire bonding technique. The cooling system uses micro-channel ceramic cooling blocks to extract a power of \SI{300}{\watt} from the scanner volume. A technique to assemble the TT-PET scanner was developed and tested. The results from the measurements on a thermo-mechanical mock-up are presented and compared with FEA simulations.}

\keywords{}

\begin{document}
\maketitle
\flushbottom

\section{The TT-PET scanner layout}
\label{sec:intro}
\subsection{Scanner architecture}
\label{sub:scanner}
\begin{figure}[htbp]
	\centering
	\includegraphics[width=0.6\textwidth]{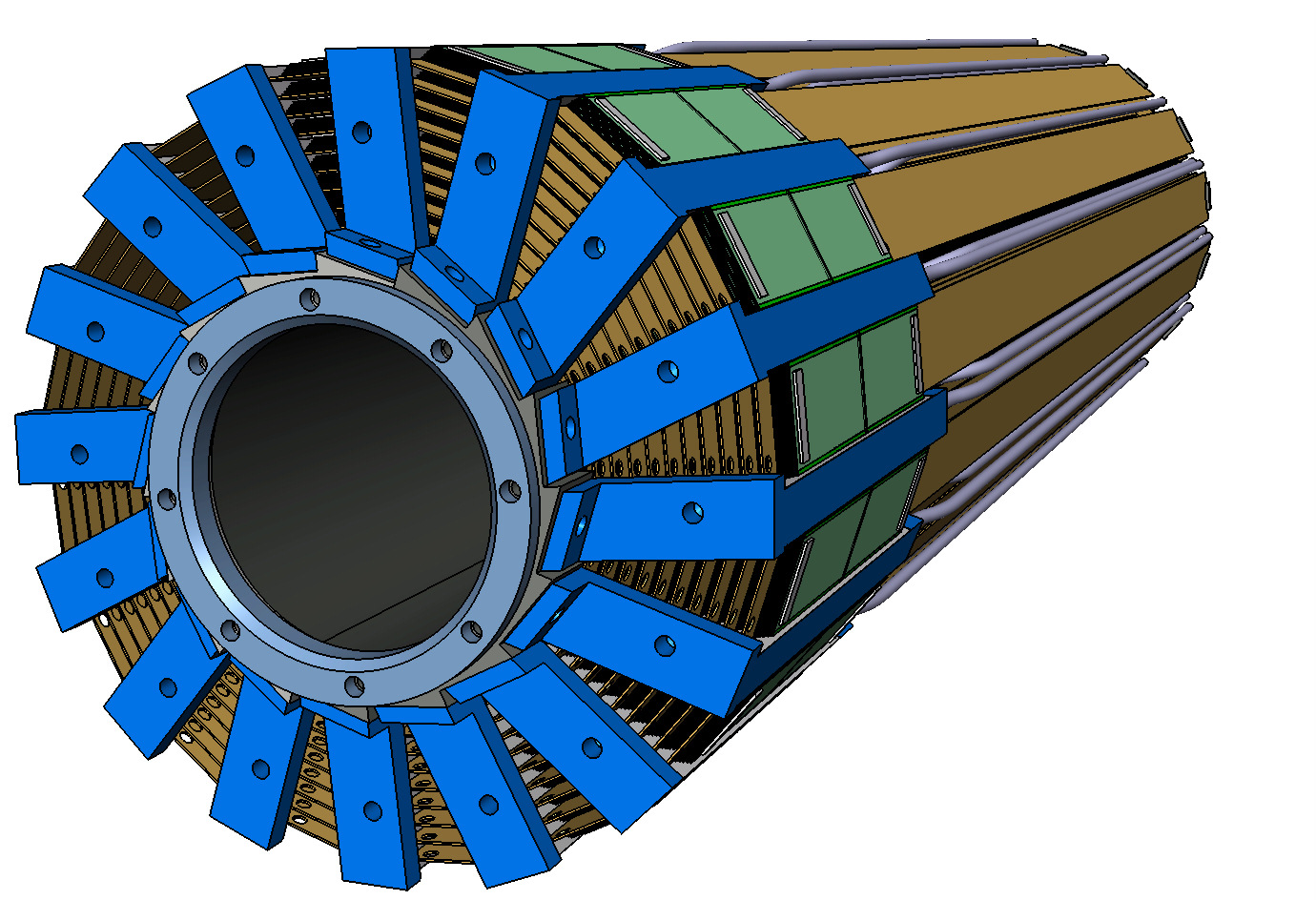}
	\caption{CAD image of the TT-PET scanner, with the 16 towers and the cooling blocks between them represented in blue. The wedge-shaped towers are formed by ASICs of three sizes, with larger ones at larger radii.}
	\label{fig:scannerlayout}
\end{figure}

The proposed small-animal TOF PET scanner (shown in figure \ref{fig:scannerlayout}) is a layered detector for \SI{511}{\kilo\electronvolt} photons based on monolithic timing pixel sensors purposely developed for this project. It will contain 1920 detector ASICs, for a total of approximately 1.5 million pixels, corresponding to a 3D granularity of 500$\times$500$\times$\SI{220}{\cubic\micro\meter}. The pixel detector has a target time resolution of $ \SI{30}{\pico\second} $ rms for the measurement of electrons over the entire detection volume, for which it requires a synchronization of all the ASICs to a \SI{10}{\pico\second} precision. The scanner is formed by 16 identical stacks of detectors, called ``towers'', shaped as wedges and interleaved by cooling blocks (described in detail in section \ref{sec:thermals}). A tower has a total thickness of \SI{22.5}{\milli\meter} and is formed by 60 detection layers, called modules, grouped in super-modules of 5 layers each.
\\

\subsection{Module layout and detection mechanism}
\label{sub:types}
\begin{figure}[htbp]
	\centering
	\includegraphics[width=0.6\textwidth]{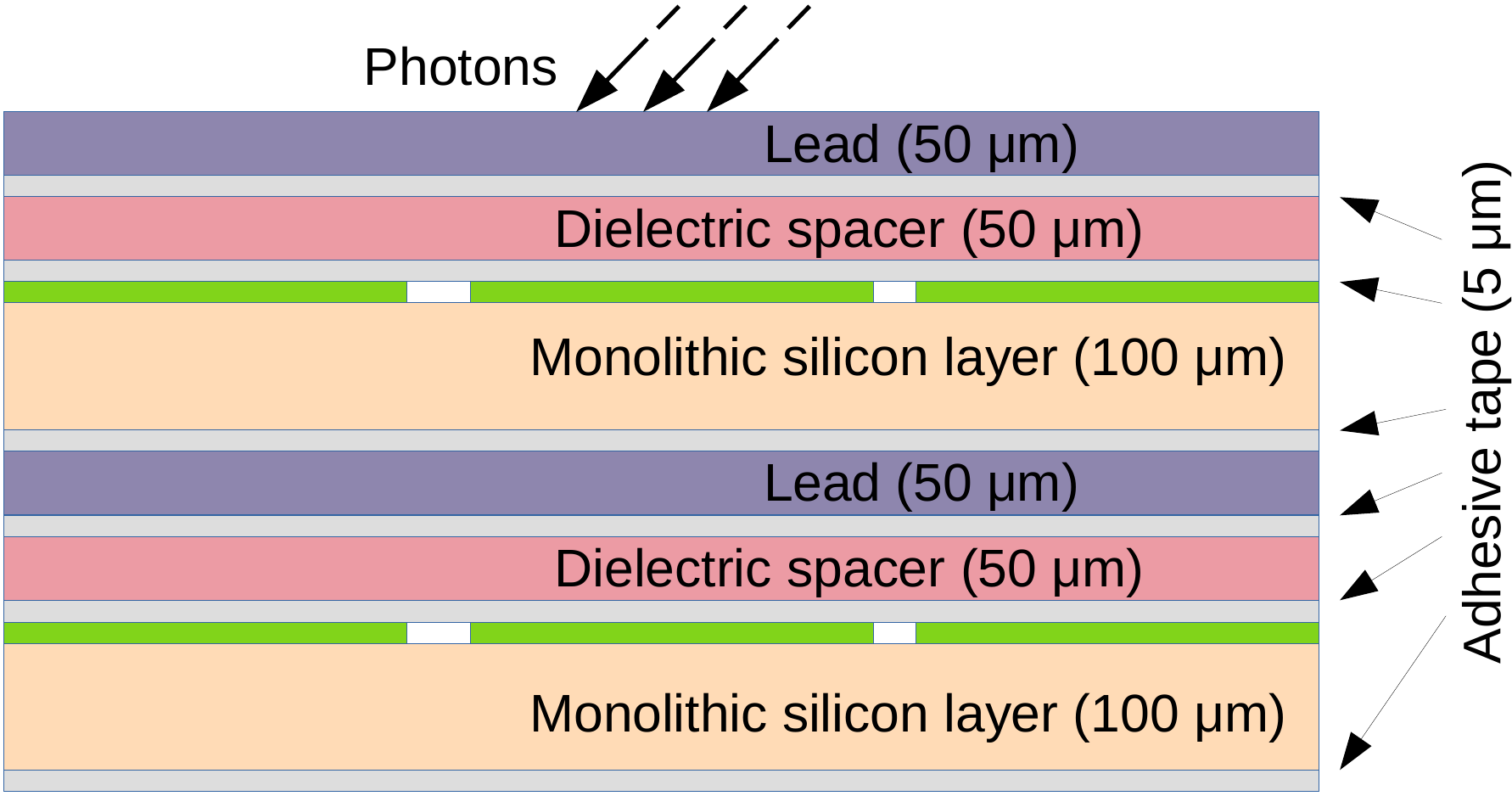}
	\caption{Two detection modules, including monolithic silicon detectors, lead converters, dielectric spacers and adhesive tape. The 60 detection modules of a tower are divided in 12 super-modules of 5 layers each. The lead and silicon layers are glued together with \SI{5}{\micro\meter} thick double-sided adhesive tape.}
	\label{fig:stack}
\end{figure}

The photon detection element of the scanner is the module, which is a layered structure made of a \SI{50}{\micro\meter} thick lead absorber, a dielectric spacer and a \SI{100}{\micro\meter} thick monolithic silicon pixel timing detector. A stack of two modules is shown in figure \ref{fig:stack}. The layers are connected to each other by means of a \SI{5}{\micro\meter} double-sided adhesive tape. The dielectric spacer has a thickness of \SI{50}{\micro\meter} and is made by a low permittivity material, necessary to reduce the capacitive coupling between the sensor pixels and the lead absorber. Inside the lead layer the photon generates a high energy electron by Compton scattering or photoelectric interaction, which traverses the dielectric spacer and deposits energy in the pixel detector. Geant4 simulations show that in almost 50\% of the cases the electron is reflected back by the following lead layer. This effect has the advantage of increasing the average signal inside the pixel, but at the risk of increasing the cluster size for electrons emitted at large angles. For this reason the distance between the ASIC backplane and the lead layer should be kept as small as possible.
\\
Three monolithic active pixel sensor types are needed to maximize the sensing area at larger radii of the tower stack-up. The modules are then grouped in three categories according to their area: 7$\times$\SI{24}{\milli\meter\squared}, 9$\times $\SI{24}{\milli\meter\squared} and 11$\times$\SI{24}{\milli\meter\squared}. The total power consumption scales linearly with the area of the module. In order to cover a larger area, two chips are used in each layer, side by side.

\subsection{The super-module}
\label{sub:supermodule}
\begin{figure}[htbp]
	\centering
	\includegraphics[width=0.9\textwidth]{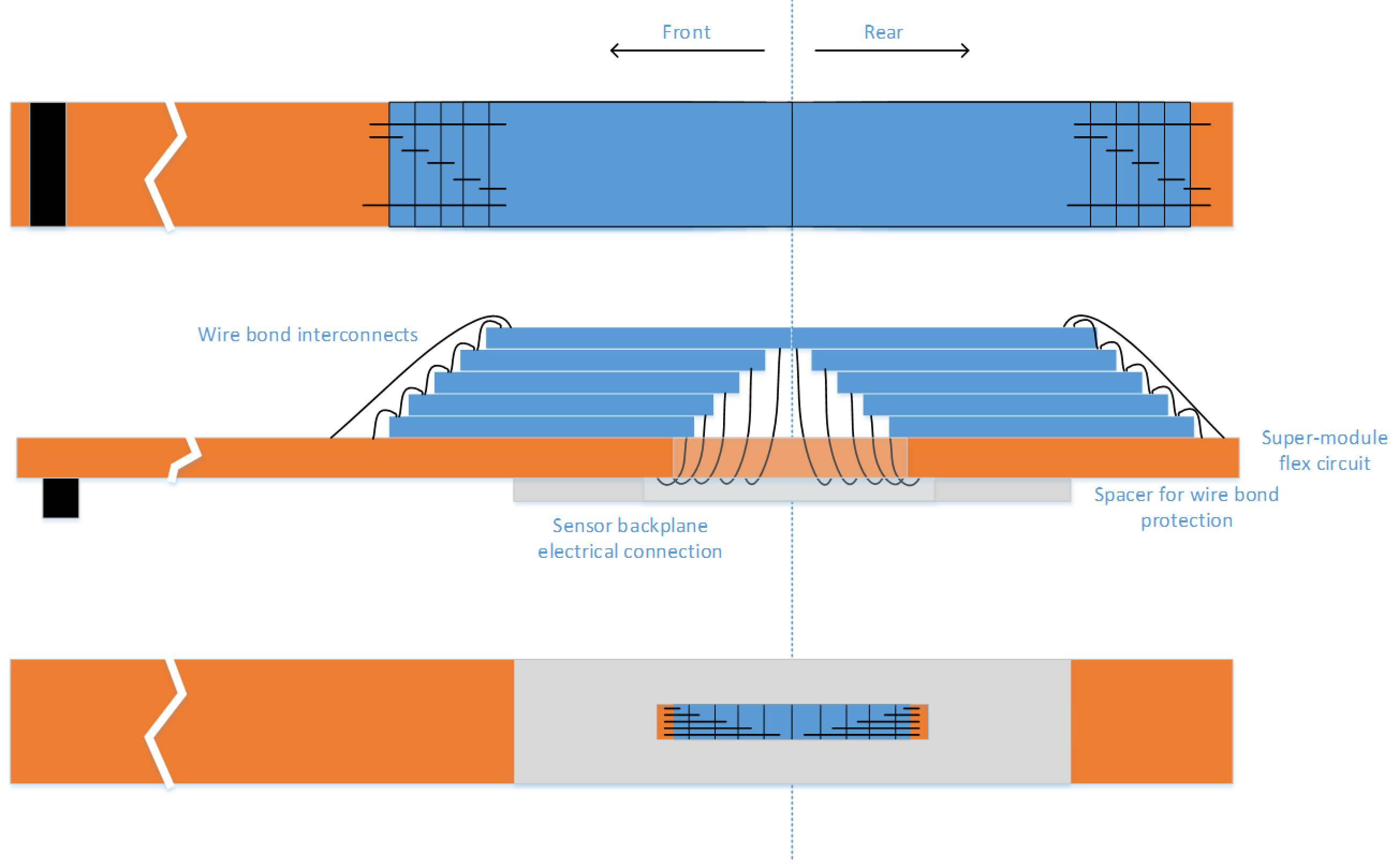}
	\caption{Super-module stack-up serviced with a flex circuit at the bottom and wire bond interconnections. In the construction, five modules are glued and staggered at the rear and front sides. A spacer with a centered slot is glued at the bottom to protect the wire bond envelopes.}
	\label{fig:wirebond-scheme}
\end{figure}

The super-module consists of an assembly of 10 modules staggered in five layers as illustrated in the central part of figure \ref{fig:wirebond-scheme} and displayed at the front and rear sides. Each module is electrically connected via wire bonds to the super-module flex PCB. The front side is the only open access for the services of the scanner. The pigtail of each super-module connects radially to a patch panel which distributes the power, clock and data lines. Each service flex bends radially when connecting to the patch panel; for this reason the tower includes a stress release mechanism in order not to transmit any force to the module stack-up and to the wire bonds located at the front side.\\
Wire-bond connections are made on both sides of the stack: on the top for the ASIC power and data signals and on the bottom to reference the sensor backplane to ground. The stacked die wire bond concept and tests are discussed in section \ref{sub:interfaces}. In the assembly sequence, the back side wire bonds are made first, then a \SI{300}{\micro\meter} thick spacer for wire bond protection is glued. The function of the latter is to protect the wire bond inside a mechanical envelope as well as to protect the top side wire bonds when assembling super-modules together.
\\
One of the challenges in the conceptual design of this detector is to keep the tower thicknesses below \SI{22}{\milli\meter}\footnote{The size of the scanner was calculated to allow its insertion in a small animal MRI machine, with the target of performing combined PET/MRI scans on mice.}. With the nominal thickness of the components, the stack-up should reach \SI{20.4}{\milli\meter} thickness. The only way to assemble such complex multilayer structure is to use a \SI{5}{\micro\meter} thick  double coated adhesive tape that should allow matching the above requirement.

\section{Electrical Interfaces}
\subsection{Power and I/O distribution}
\label{sub:interfaces}
The challenge of providing interconnections to all the chips in a super-module imposed the development of a specific communication protocol that allows the 10 ASICs to be connected in a daisy-chain and behave as a single larger chip. In this way, most of the signals can be provided to the super-module with a single line, being then shared or connected only to the first chip in the chain and then propagated from chip to chip.
The only exception to this strategy is the global synchronization signal: while it is logically the same signal being distributed to all chips, having a low-jitter line is very important to the chip performance, as any uncertainty on this signal is directly added to the time resolution of the system. It was thus chosen to provide a separate differential synchronization line to each chip. Table \ref{tab:signals} lists the signals needed for each ASIC.
\begin{table}[htbp]
\centering
\begin{tabular}{|c|c|c|}
	\hline 
	\textbf{Signal} & \textbf{Differential} & \textbf{Type of connection} \\ 
	\hline 
	Analog VDD & No & Shared \\ 
	\hline 
	Digital VDD & No & Shared \\ 
	\hline 
	Analog Gnd & No & Shared \\ 
	\hline
	Digital Gnd & No & Shared \\ 
	\hline 
	HV & No & Shared \\ 
	\hline
	Guard ring bias & No & Shared \\ 
	\hline
	Sync & Yes & Point-to-point \\ 
	\hline 
	MOSI & Yes & Daisy-chained \\ 
	\hline 
	MISO & Yes & Daisy-chained \\ 
	\hline
	Clock out & Yes & Daisy-chained \\ 
	\hline
	Trigger & Yes & Daisy-chained \\ 
	\hline
	Reset & No & Shared \\ 
	\hline
\end{tabular} 
\caption{List of signals that the super-module needs to operate and communicate to the readout system. Daisy chaining lines from chip to chip means that it is necessary to connect chips physically stacked on top of each other.}
\label{tab:signals}
\end{table}
 
\subsection{Dummy super-module construction and wire bonding tests}
\label{sub:dummy}
\begin{figure}[htbp]
	\centering
	\begin{minipage}{0.45\textwidth}
		\centering
		\includegraphics[width=0.9\textwidth]{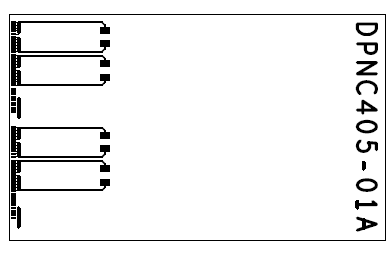}
	\end{minipage}\hfill
	\begin{minipage}{0.55\textwidth}
		\centering
		\includegraphics[width=0.9\textwidth]{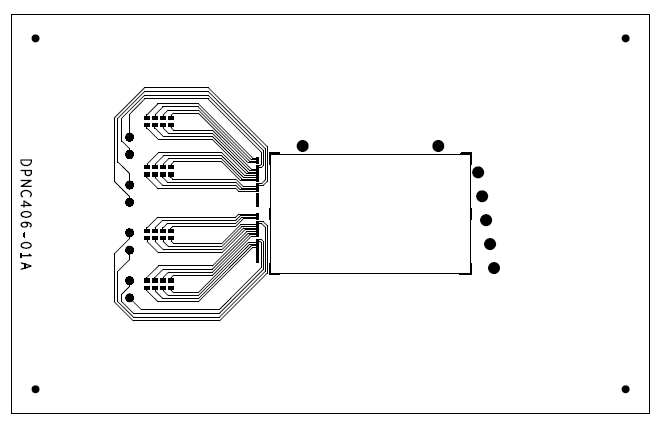}
	\end{minipage}
	\centering
	\caption{Layout of the PCBs used for wire bonding tests. On the left, the smaller PCB simulating the ASIC, on the right the bigger one hosting the stack to be tested that emulated the sumer-module flex.}
	\label{fig:pcb-layout}
\end{figure}

Two PCBs (shown in figure \ref{fig:pcb-layout}) were designed and produced with a standard glass-reinforced epoxy laminate substrate, to study the feasibility of the electrical interconnection between the modules and the upstream services. The first one, simulating the super-module flex connected the read-out electronics, was designed with a resistor network to check the integrity of the circuitry after wire bonding. The second dummy board was designed to mimic the module bond-pad interconnections. It has a \textasciitilde\SI{200}{\micro\meter} thickness and a similar surface area to the ASIC. An additional requirement of the final stack-up is an adhesive layer between modules offering a small and uniform thickness. The \SI{5}{\micro\meter} double-coated adhesive tape used for this test was originally developed for compact electronics applications such as cellular phone and digital cameras. The adhesion performance was found to be exceptionally good in terms of shear strength while remaining acceptable with respect to the peeling characteristics. A first test showed how the wire bonding process often failed in the presence of a small bow of the glued substrates, as the bow was causing the edge of the dummy chips to lift by a few microns.\\
The application technique and sequence were optimized in order to obtain a reliable adhesion performance of the tape, as the procedure proved very sensitive to them. With the new procedure the adhesion was very consistent for glass, plastic and low-surface-roughness metal samples. No visible trapped air bubble was detected and the measured thickness was uniform within \SI{1}{\micro\meter}.
\\
Independent measurements were made to assess the resistance to shear stress and all the values consistently exceeded the specifications given by the manufacturer of \SI{265}{\newton} for an adhesion surface of 20$\times$\SI{20}{\milli\meter\squared} at room temperature. Thermal cycling tests were also performed between 10 and \SI{50}{\celsius} without any visible change, which is consistent with the technical datasheet for which adhesion performance was given for a range of 10 to \SI{80}{\celsius}.\\
The staggered assembly of the five dummy modules was made with a simple jig with alignment pins. The step depth was designed to be \SI{500}{\micro\meter}, which is considered a realistic target for the wire bonding access in between the layers. The thin tape was first attached to each module and then the liner was removed only when connecting the modules to the parts already assembled. The modules were handled with a vacuum tool and the application force was applied according to the successfully tested procedure. The wire-bonding was then made targeting a maximum loop height of \SI{200}{\micro\meter} and a maximum and uniform pull strength of more than \SI{10}{g}. The measurements on the stack-up showed a maximum loop height of about \SI{70}{\micro\meter} and an average pull strength of \SI{12}{g} with less than 5\% dispersion.\\
A wire-bond routine was made to test the reliability of the process. Two samples were wire bonded without any failure and completed in less than 15 minutes each. The samples were then cycled ten times in temperature inside a climate chamber from 0 to \SI{50}{\celsius} and the electrical continuity was successfully checked with a resistor network and probe pads on the larger PCB.\\
A protection technique consisting of spraying the wire bonds with polyurethane in order to protect them against possible chemical issues due to water contamination or non-intentional mechanical stresses was also used\footnote{ The same technique was tested for the Insertable Barrel Layer of the ATLAS experiment at CERN\cite{IBL}.}. After this procedure, additional thermal cycling tests were performed, without showing any measurable degradation. 
\begin{figure}[htbp]
	\centering
	\begin{minipage}{0.35\textwidth}
		\centering
		\includegraphics[width=0.9\textwidth]{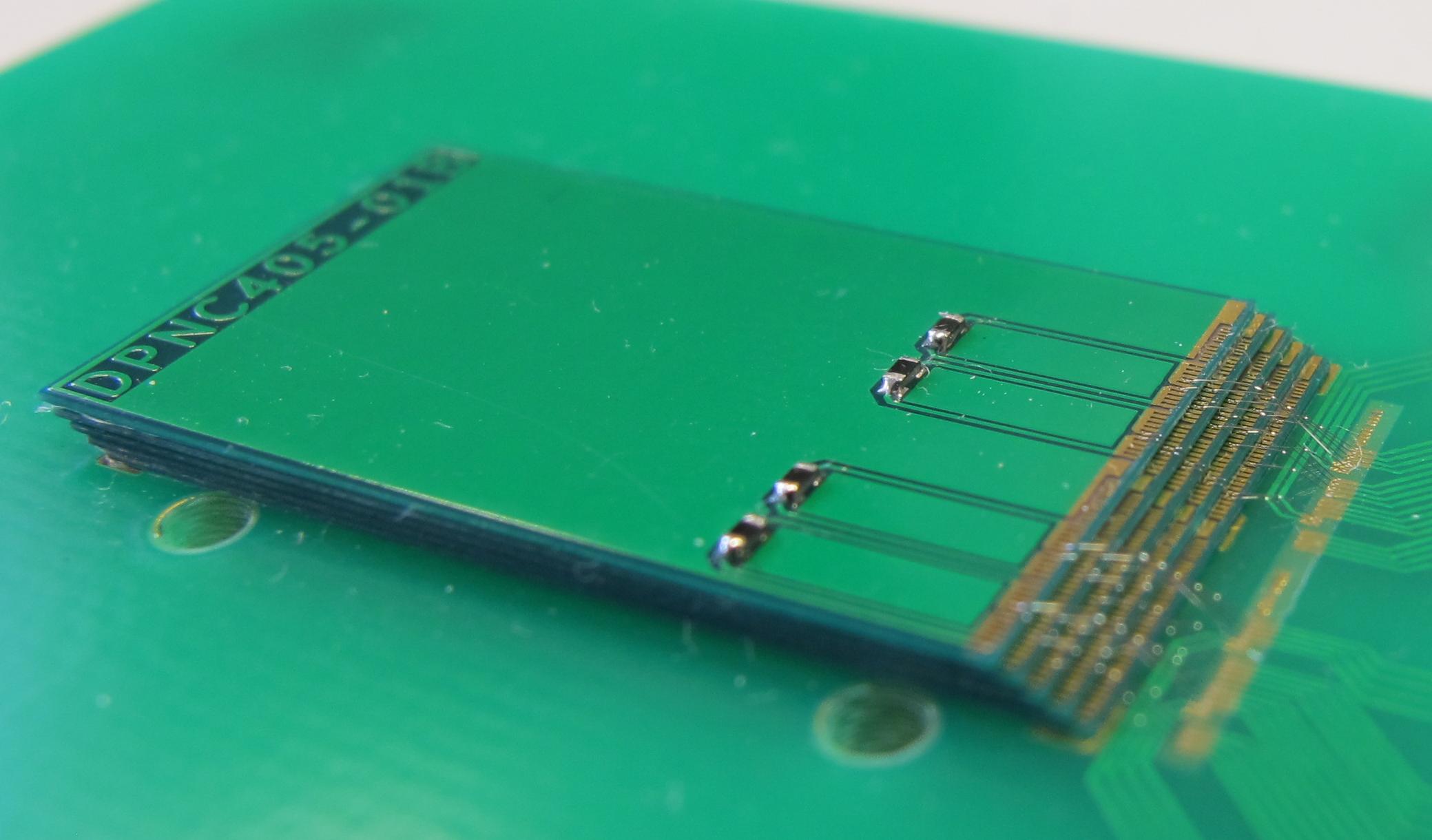}
	\end{minipage}\hfill
	\begin{minipage}{0.35\textwidth}
		\centering
		\includegraphics[width=0.9\textwidth]{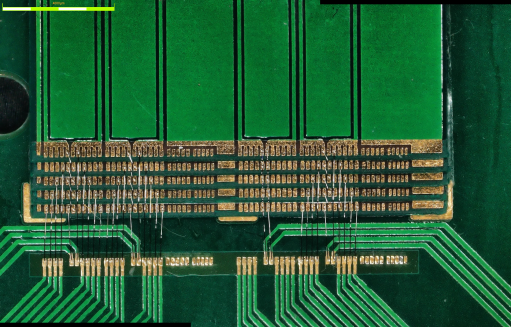}
	\end{minipage}\hfill
	\begin{minipage}{0.25\textwidth}
		\centering
		\includegraphics[width=0.9\textwidth]{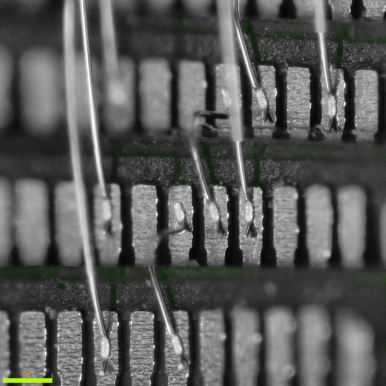}
	\end{minipage}
	\centering
	\caption{Pictures of the stacked wire-bond test. The left picture shows the staggered assembly glued together using ultra-thin double coated tape. The central picture illustrates the wire bond made over the six pad layers. The right picture was taken with a microscope showing the bond foot of three layers.}
	\label{fig:wirebond-pics}
\end{figure}\\

\section{Thermal management}
\label{sec:thermals}

\subsection{Requirements and cooling strategy}
\label{sub:requirements}
\begin{figure}[htbp]
	\centering
	\includegraphics[width=0.9\textwidth]{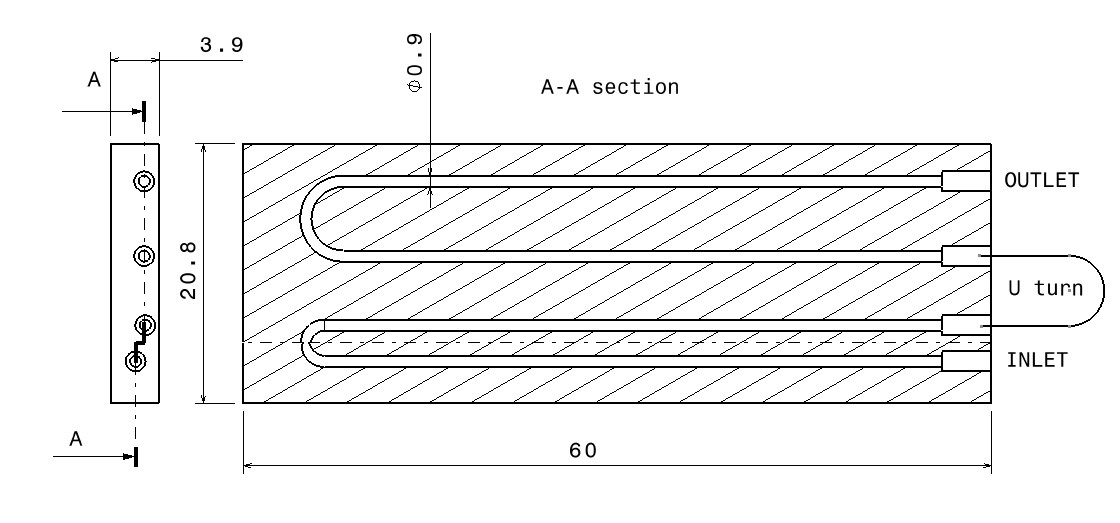}
	\caption{Drawing of the cooling block for the thermal mock-up. Dimensions are in \SI{}{mm}.}
	\label{fig:coolingdrawing}
\end{figure}

\begin{figure}[htbp]
	\centering
	\begin{minipage}{0.2\textwidth}
		\centering
		\includegraphics[width=0.9\textwidth]{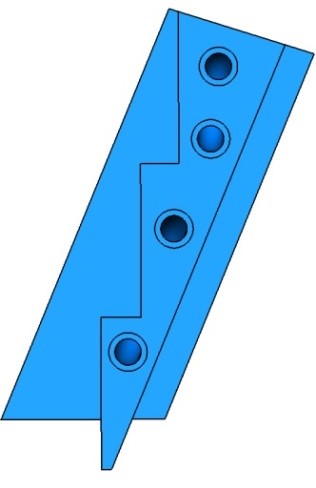}
	\end{minipage}\hfill
	\begin{minipage}{0.8\textwidth}
		\centering
		\includegraphics[width=0.9\textwidth]{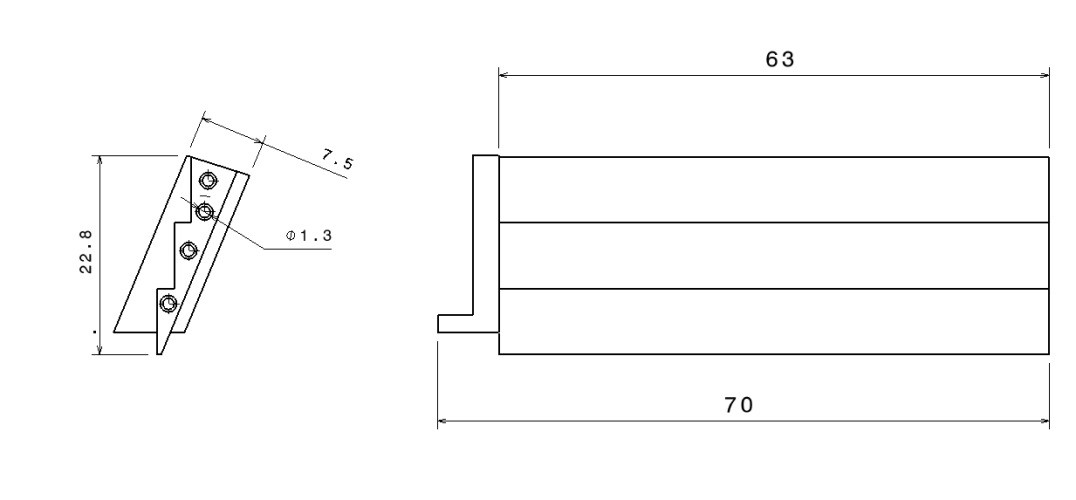}
	\end{minipage}
	\centering
	\caption{Cross section of the modified cooling block proposed for the final design, with thicker holes compared to the measured prototype.}
	\label{fig:newsize}
\end{figure}

\begin{figure}[htbp]
	\centering
	\includegraphics[width=0.5\textwidth]{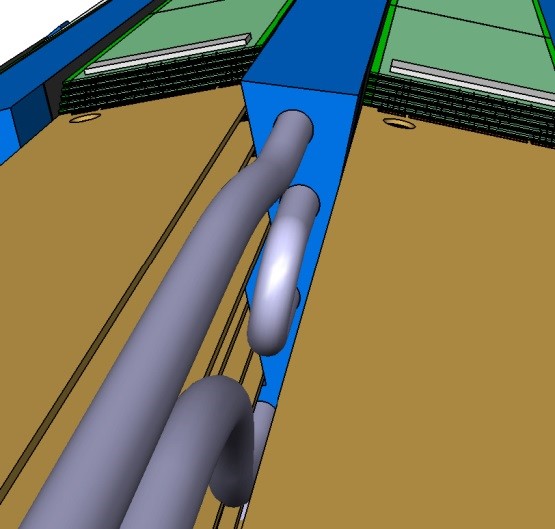}
	\caption{CAD drawing of the inlet/outlet cooling block pipes.}
	\label{fig:pipes}
\end{figure}

Each TT-PET monolithic chip features many detection channels, that are required to be powered at all times. While the front-end uses a relatively low power compared to similar detectors (less than \SI{80}{\milli\watt\per\square\centi\meter}, see \cite{pierpaolo} and \cite{ivan}), this number is still much higher than the power consumption of any other block inside the ASIC (TDC, readout logic...) due to the high granularity of the sensor. We can thus assume that all the power is drawn by the pixel front-ends, placed in two rows at the long edges of the chips.\\
In terms of thermal management of the detector the key items in the design are the cooling blocks that are interposed between the towers. The total power to be dissipated is not very high (\textasciitilde\SI{300}{\watt}) but is contained in a small and confined volume. Therefore, particular attention has to be taken to ensure sufficient flow for the cooling fluid. The cooling block has to extract the heat through the fluid circulating across the channels. The challenge is to identify a non-metallic material that can be built with micro-channeling circuitry, which could be interfaced to the outside world through pipes and manifold systems. The target minimum Heat Transfer Coefficient (HTC) for the cooling blocks estimated from FEA simulations is \SI{4000}{\watt\per\meter\squared\per\kelvin}.\\
In order to qualify possible materials and designs, prototypes of the cooling blocks were built and later used in a thermal mock-up of the scanner tower.
\begin{figure}[htbp]
	\centering
	\begin{minipage}{0.35\textwidth}
		\centering
		\includegraphics[width=0.9\textwidth]{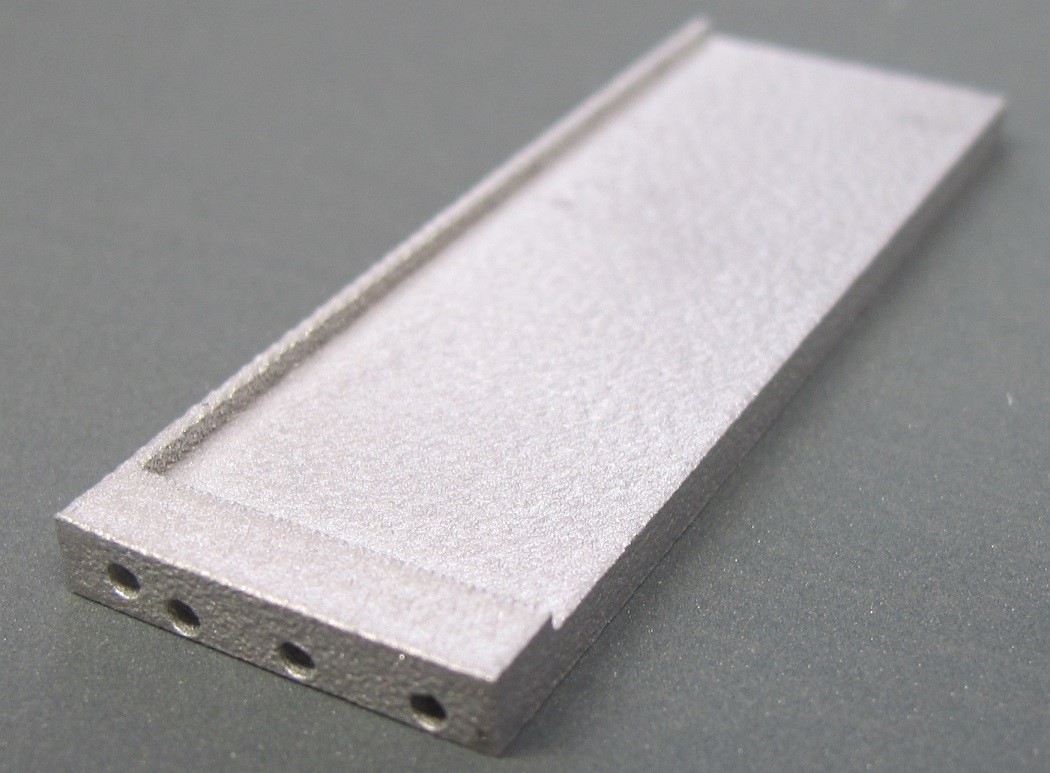}
	\end{minipage}\hfill
	\begin{minipage}{0.65\textwidth}
		\centering
		\includegraphics[width=0.9\textwidth]{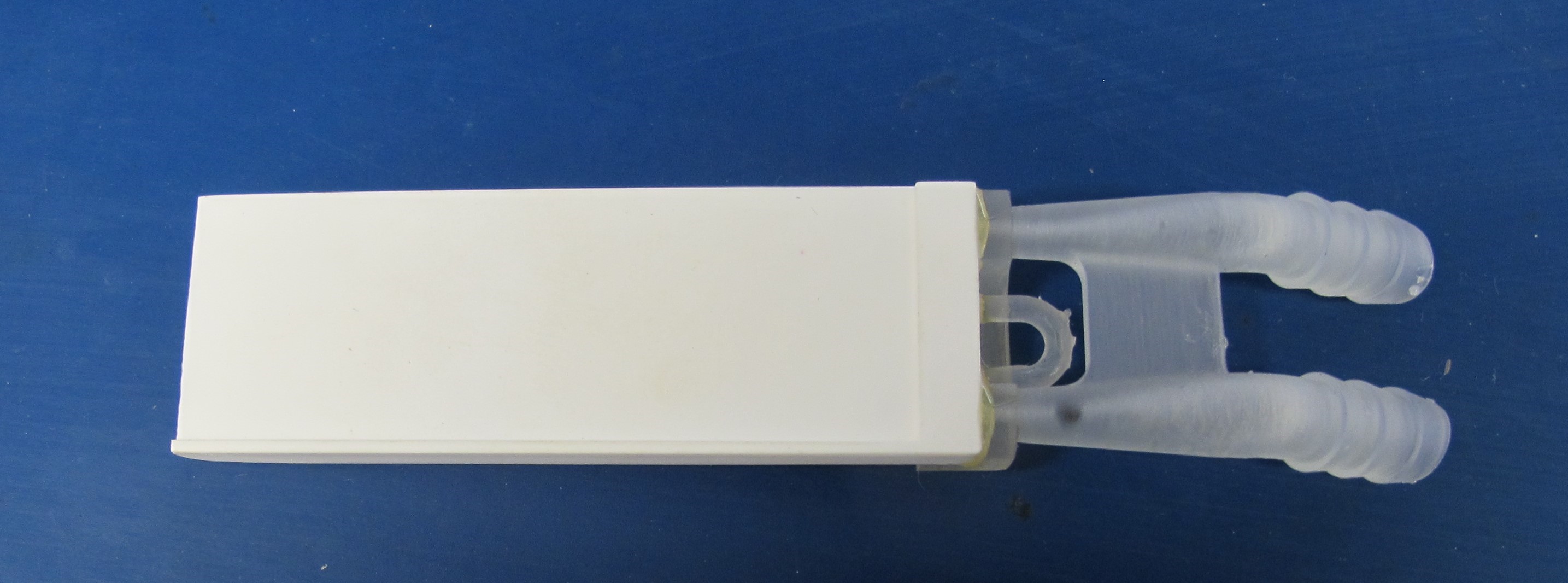}
	\end{minipage}
	\centering
	\caption{Two types of cooling blocks were produced based upon the same design. One made of aluminum (on the left) and the second made of aluminum oxide ceramic (on the right).}
	\label{fig:coolingblock}
\end{figure}\\
The wedge structure of the tower was simplified for the cooling block prototypes using a uniform thickness along the block length and width. The two types of blocks, shown in figure \ref{fig:coolingblock}. one made of aluminum and another made of aluminum oxide ceramic (\ce{AlO3}), were fabricated by laser sintering method to assess their feasibility, performance and reliability. The typical precision that is achievable with laser sintering is in the order of \SI{100}{\micro\meter} in all directions, and the minimum wall thickness guaranteed by manufacturer is \SI{400}{\micro\meter}. The design of both the thermal mock-up (see figure \ref{fig:coolingdrawing}) and the final version is done according to the aforementioned constraints. The process was found to be feasible in terms of manufacturability, but offered a poor yield. Out of the six parts produced two were showing an excellent coolant flow while the other four failed due to partial or full clogging. The main reason is that the ratio between the channel length and its diameter is such that the channels cannot be evacuated easily during the construction and residual material tends to clog the pipes after a while. The final design will be modified so that the diameter of the hole will increase by about 50\% while the wall thickness will be kept the same.
\\
In the thermal mock-up the inlet/outlet fittings and U turns were 3D-printed with clear resin in stereo lithography techniques. This is a solution to investigate for the final version, whereas the access and spac between towers are very limited (standard flexible pipes in the lowest radii might interfere with module flexes).\\
As a result of this test, the diameter of the final cooling block (see figure \ref{fig:newsize}) is foreseen to be \SI{1.3}{\milli\meter} and the inter distance between channels is balanced over the block height. This will limit the effect of channel clogging. To reduce the risk even more, the cooling block will be divided in 3 parts: the central and ``linear'' segment (only straight channels to ease the ``cleaning'' after processing), and 2 extremities that will be glued and sealed together. Figure \ref{fig:pipes} shows the limited space available at small radius in the tower assembly. A design currently under study foresees to produce also the extremities with laser sintering, integrating the needed channel U turns and organizing the inlet/outlet fitting at higher radius to avoid any interference with the module flexes.\\
\\
The manifolding between towers will be organized outside the detector envelope, so that only 2 pipes will get in and out from the patch panel with flexible tubing. More detailed studies will be done on the connections (fitting parts) of the cooling blocks to allow for modularity and ease of maintenance.
\\

\subsection{Cooling test on the scanner thermal mock-up}
\label{sub:mockup}
\begin{figure}[htbp]
	\centering
	\begin{minipage}{0.45\textwidth}
		\centering
		\includegraphics[width=0.9\textwidth]{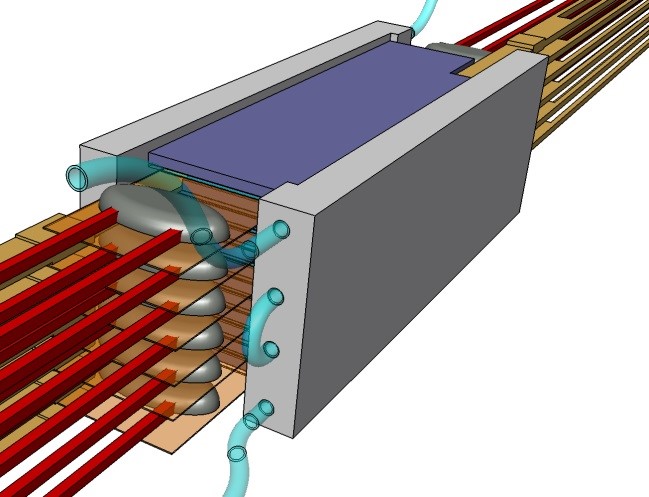}
	\end{minipage}\hfill
	\begin{minipage}{0.55\textwidth}
		\centering
		\includegraphics[width=0.9\textwidth]{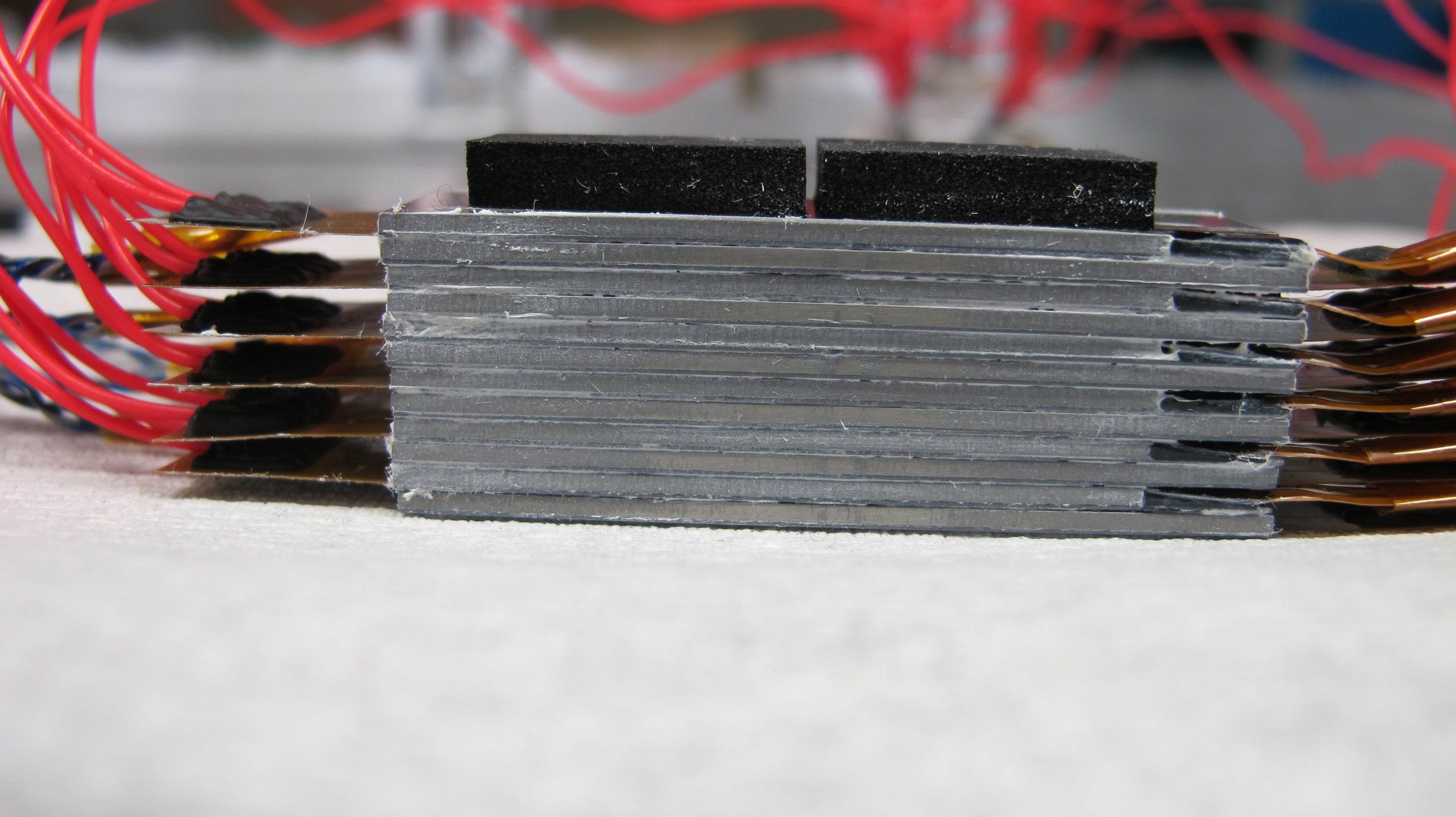}
	\end{minipage}
	\centering
	\caption{On the left is a CAD view of the thermal super-module with the cooling block attached and the pipe interconnection represented in translucent color. On the right is a picture of the thermal mock-up showing services of the odd and even super-module exiting on the two sides.}
	\label{fig:thermal}
\end{figure}

\begin{figure}[htbp]
	\centering
	\includegraphics[width=0.7\textwidth]{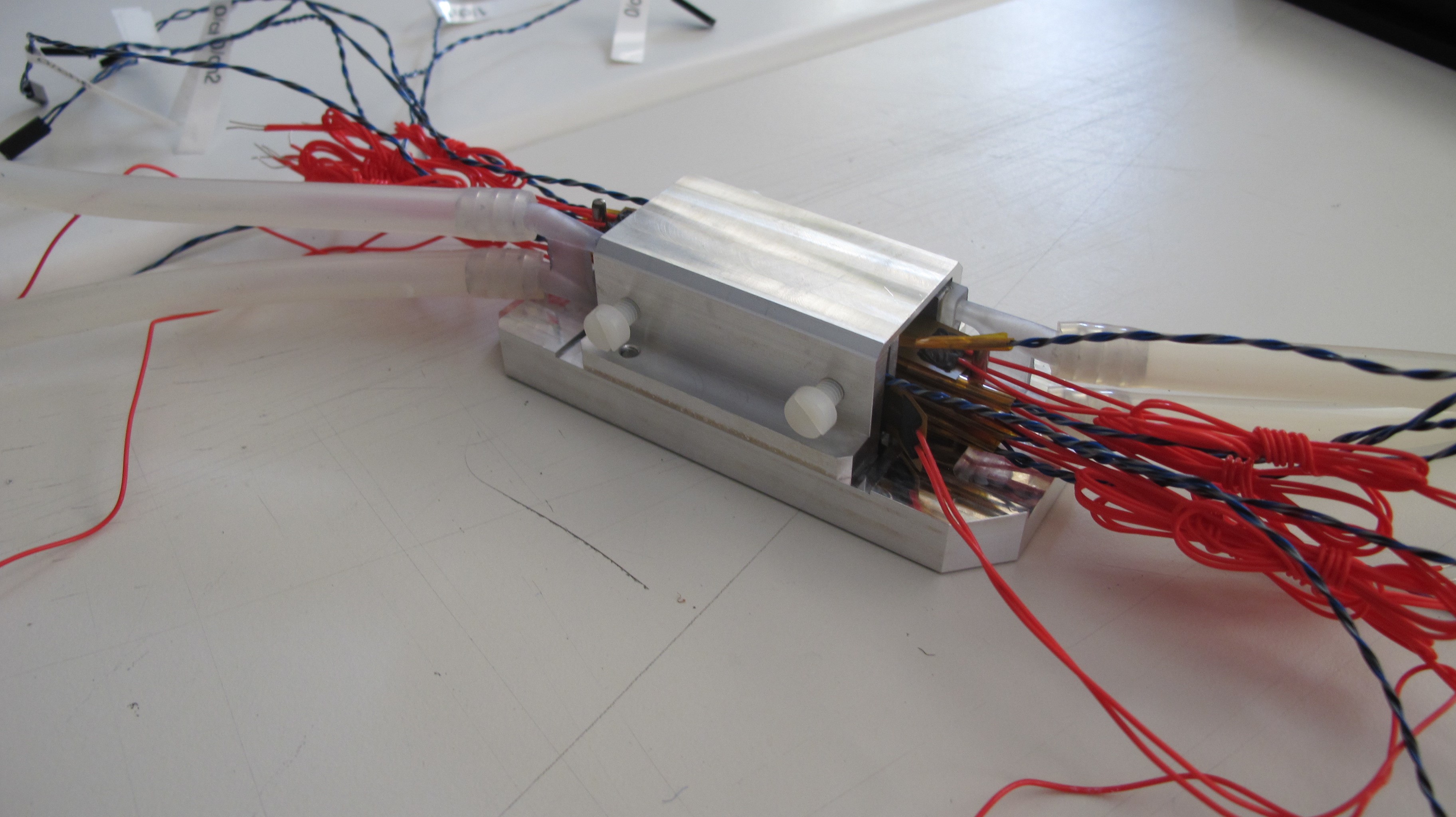}
	\caption{Thermal mock-up with cooling blocks. The red wires are servicing the heater pads which are segmented into four power supply channels. The inlet and exhaust tubes are split in order to circulate the cooling fluid in parallel into the two blocks.}
	\label{fig:thermal-setup}
\end{figure}

A thermal mock-up of the tower (shown in figure \ref{fig:thermal}) was built to demonstrate the feasibility of the tower assembly with the interface to the cooling block, and to verify that the operating temperature remains below \SI{40}{\celsius}. The mock-up was connected to a chilled water system with a regulated inlet temperature at about \SI{10}{\celsius} and no pipe insulation installed along the distribution lines. In this condition a relative humidity of less than 50\% at an ambient temperature of \SI{21}{\celsius} is required in order to stay below the dew point.
The integrated electronics does not have strict temperature constraints, but the assembly must remain mechanically stable with minimum built-in stresses due to CTE mismatch. The total power, estimated to be \SI{18}{\watt} per tower, requires to have a cooling capacity of \SI{320}{\watt} including 10\% coming from the service power losses.\\
The thermal mock-up was equipped with heater pads and temperature sensors. The tower consisted of 12 super-modules each composed of 5 parts with a maximum area of 51$\times$\SI{15}{\milli\meter\squared}. Each super-module includes:
\begin{itemize}
	\item 1 laser-cut bare \SI{300}{\micro\meter} thick silicon die
	\item 1 stainless steel water jet cut with a thickness of \SI{1}{\milli\meter}
	\item 1 thermo-foil heater pad capable to deliver up to \SI{20}{\watt} with an intrinsic resistance of \textasciitilde\SI{40}{\ohm} and with a maximum thickness of \SI{250}{\micro\meter}
	\item 1 thin film NTC thermistor with a maximum thickness of \SI{500}{\micro\meter}
	\item 3 layers of ultra-thin double side adhesive tape of \SI{5}{\micro\meter} thickness
\end{itemize}
The overall thermal super-module thickness is nominally \SI{1.565}{\milli\meter}, not far from the target value for the real super-module, and it is made of materials with very similar thermal conductivities. An FEA model was built with this thermal mock-up to make comparisons with the measurements. In order to service the super-module tower the orientation of the odd and even super-module services was alternated given the inflation of thicknesses at the location of the soldering pads transition between the kapton and the wires as illustrated in figure \ref{fig:thermal}.
\\
The cooling blocks were thermally bonded at the side of the stack-up with a thermal paste interface and held with two nylon screws on each side (see figure \ref{fig:thermal-setup}). The system was serviced with flexible silicon pipes that minimize the stress at the stereo-lithographic fittings of the cooling blocks. The cooling was set-up with a manifolding system to have parallel flow in the two blocks.
In addition, each cooling block was equipped with a heater pad allowing the injection of the equivalent power coming from the neighboring tower and with an NTC temperature sensor. The readout system of the NTC temperature sensors uses custom CAN controller readout boards. Each of the boards can handle up to 12 ADC channels, so two readout boards were daisy chained via the CAN interface links. Data were transferred to the PC via a CAN to USB interface. The system was designed to have a resolution of 0.1°C. In order to avoid systematic errors and reach the desired accuracy, each of the sensors connected to the board was calibrated for temperatures in a range between 5 and \SI{40}{\celsius} with a step of \SI{2}{\celsius}. The linearity error in this range was found to be less than 2\%.
\\

\subsection{Test results and comparison with FEA simulations}
\label{sub:results}
\begin{figure}[htbp]
	\centering
	\begin{minipage}{0.45\textwidth}
		\centering
		\includegraphics[width=0.9\textwidth]{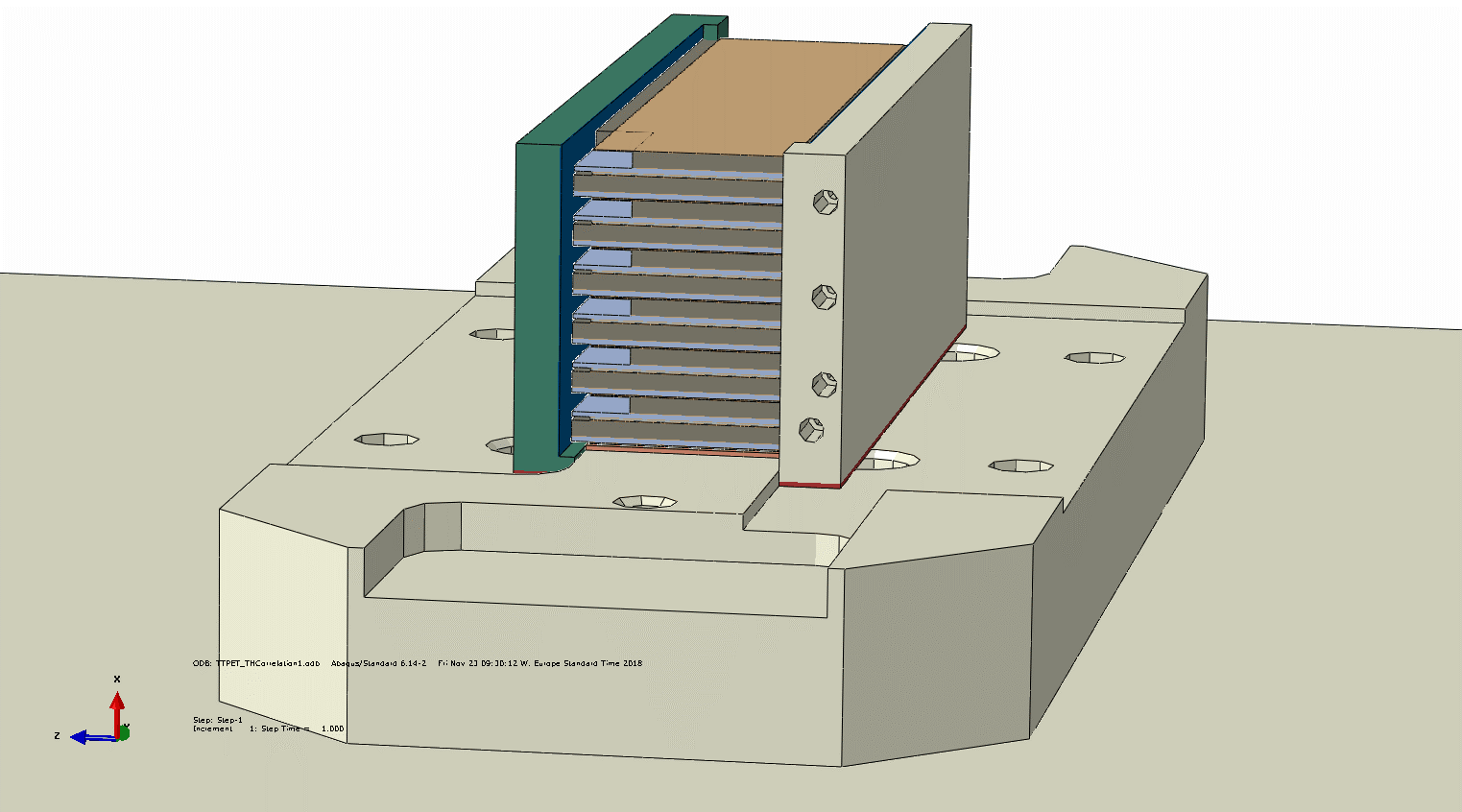}
	\end{minipage}\hfill
	\begin{minipage}{0.55\textwidth}
		\centering
		\includegraphics[width=0.9\textwidth]{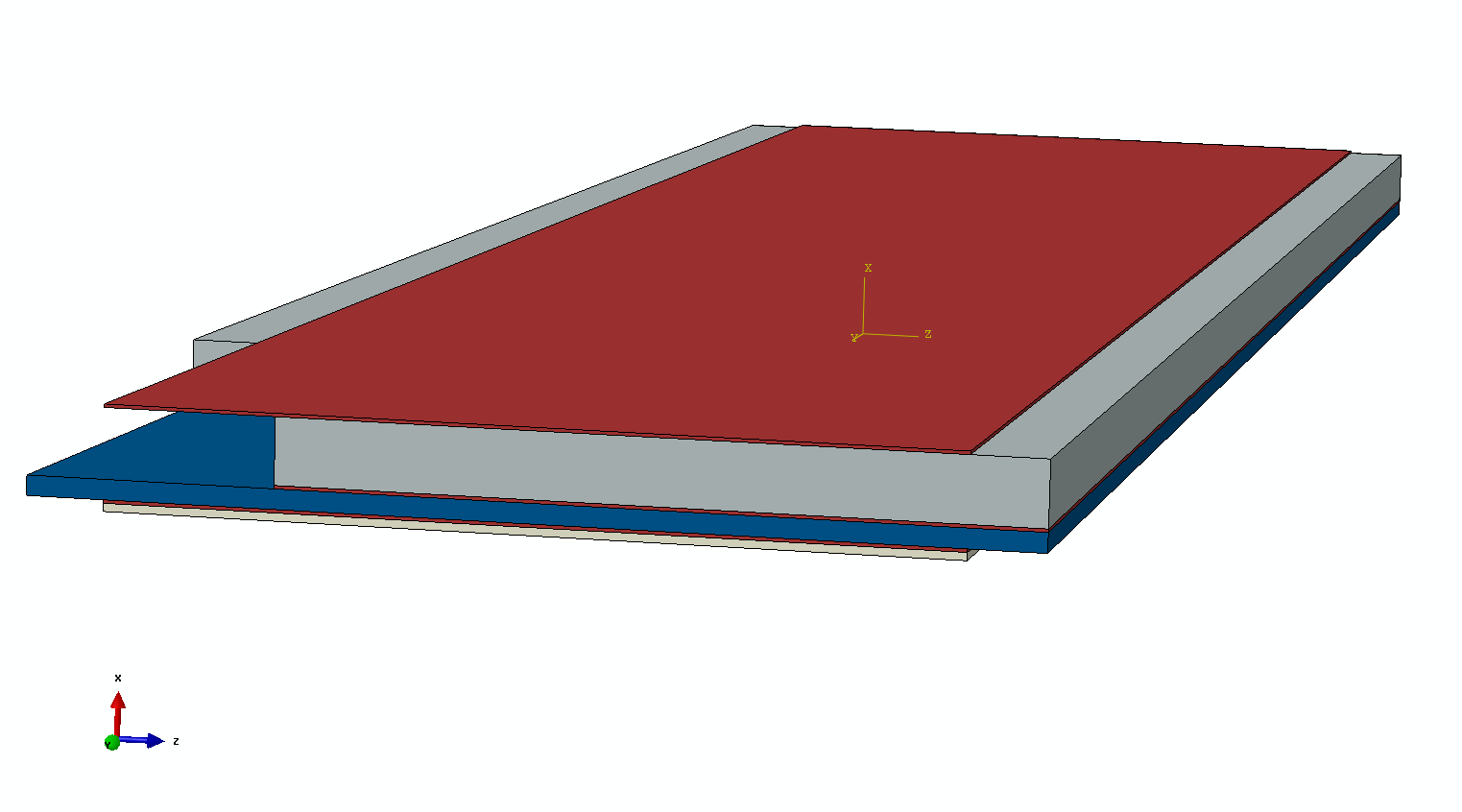}
	\end{minipage}
	\centering
	\caption{Assembly model of the full system (left) and sub-model of a super-module stack (right). The silicon plate is shown in blue, while the steel plate in gray. The heater pad where the heat flux is applied is at the very bottom.}
	\label{fig:assembly}
\end{figure}

\begin{figure}[htbp]
	\centering
	\begin{minipage}{0.5\textwidth}
		\centering
		\includegraphics[width=0.9\textwidth]{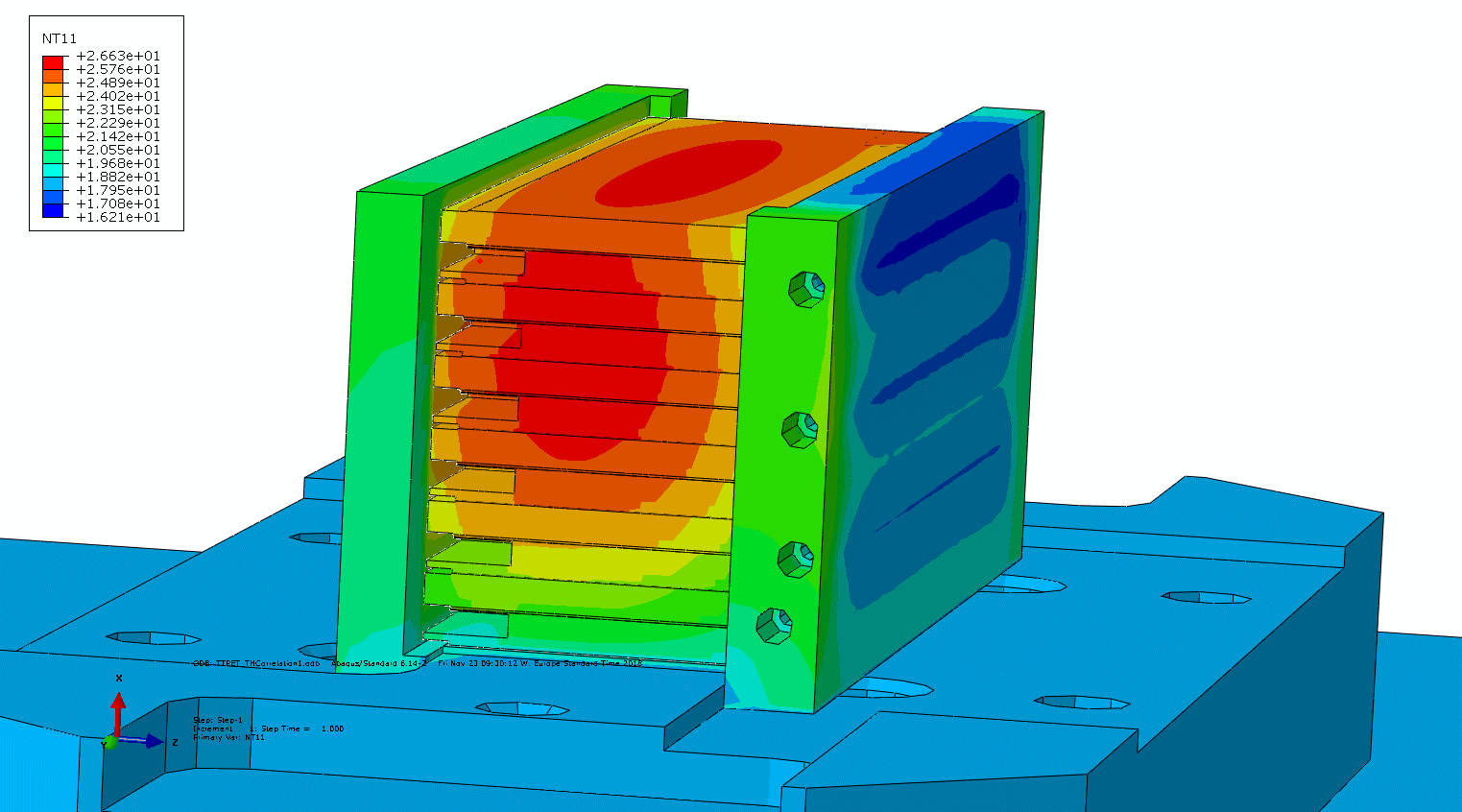}
	\end{minipage}\hfill
	\begin{minipage}{0.5\textwidth}
		\centering
		\includegraphics[width=0.9\textwidth]{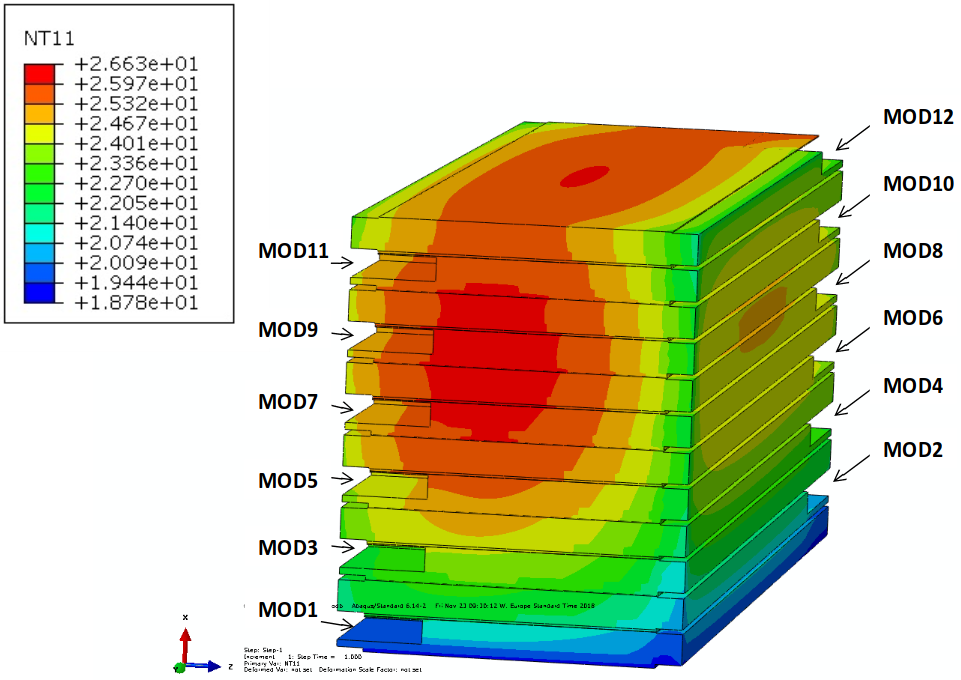}
	\end{minipage}
	\centering
	\caption{Temperature mapping of the system after HTC fine tuning. On the left, the entire block is depicted, while on the right only the module is shown. The temperatures ranges from \SI{19}{\celsius} to \SI{25}{\celsius} at NTC locations (in the voids between modules).}
	\label{fig:temperaturemapping}
\end{figure}

The flow speed of the coolant has a direct impact on the HTC and an asymmetry between the two blocks can lead to an undesired temperature distribution. Therefore, due to the yield issues described in section \ref{sub:requirements}, for the measurement campaign the mock-up was connected to the two cooling blocks with the best coolant flow, one aluminum block on one side and a \ce{AlO3} ceramic one on the other side. The total power of up to \SI{36}{\watt} was injected in 4 different points:
\begin{itemize}
	\item \SI{18}{\watt} distributed uniformly to the two heater pads of the cooling blocks
	\item \SI{6}{\watt} distributed uniformly to the 4 super-modules of the lower group 
	\item \SI{6}{\watt} distributed uniformly to the 4 super-modules of the middle group
	\item \SI{6}{\watt} distributed uniformly to the 4 super-modules of the upper group
\end{itemize}
In this set-up, thanks to parts made in stereo lithography, it was possible to precisely monitor the temperature of the coolant at the inlet and outlet of the manifold by inserting and gluing an NTC film sensor inside a little slit, directly in contact with the cooling fluid. This feature allows measuring the power exchanged between the set-up and the environment through the aluminum jig and base plate. This power was measured to be approximately \SI{6}{\watt}, a sixth of the total injected power. Moreover, even if the coolant temperature was set at \SI{10}{\celsius} there was an increase in temperature of \SI{1.8}{\celsius} due to the relatively long flexible pipe (\textasciitilde\SI{4}{\meter} back and forth). The HTC was estimated by measuring the temperature difference when no power was injected into the setup and when the full power was set, normalizing the result by the total surface between the cooling fluid and the block participating in the thermal exchange. Since it was not possible to measure the coolant temperature independently at the outlet of the two cooling blocks, the HTC was obtained averaging the heat exchange for the two types of blocks in such condition and found to be \textasciitilde\SI{8000}{\watt\per\meter\squared\per\kelvin}. This value together with the environmental temperature was then used in a thermal FEA simulation to validate the thermal model.
\begin{figure}[htbp]
	\centering
	\includegraphics[width=0.7\textwidth]{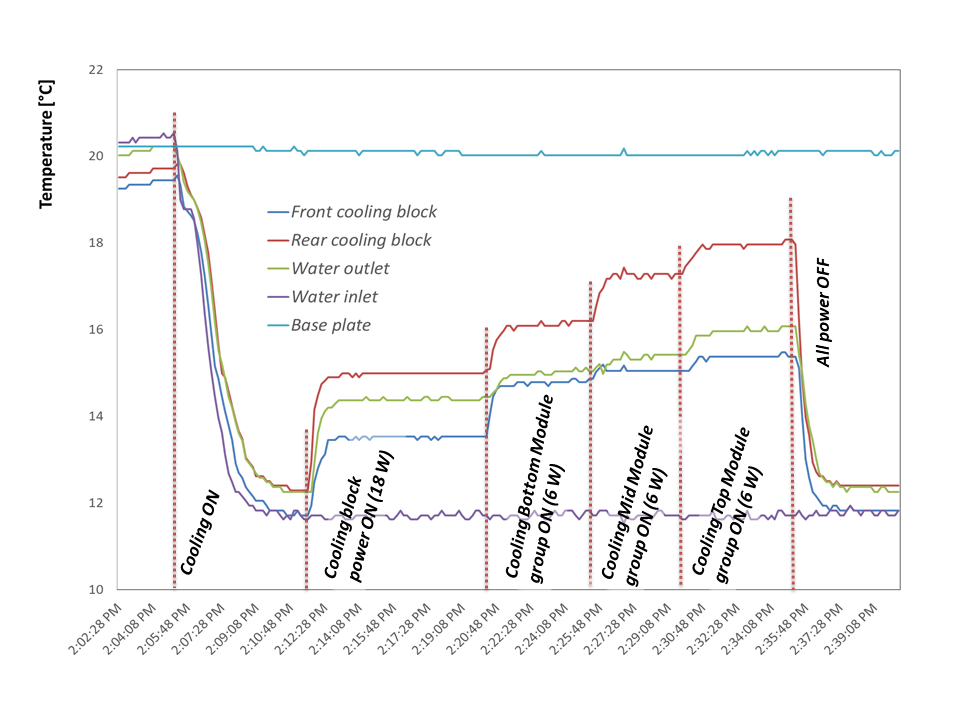}
	\includegraphics[width=0.7\textwidth]{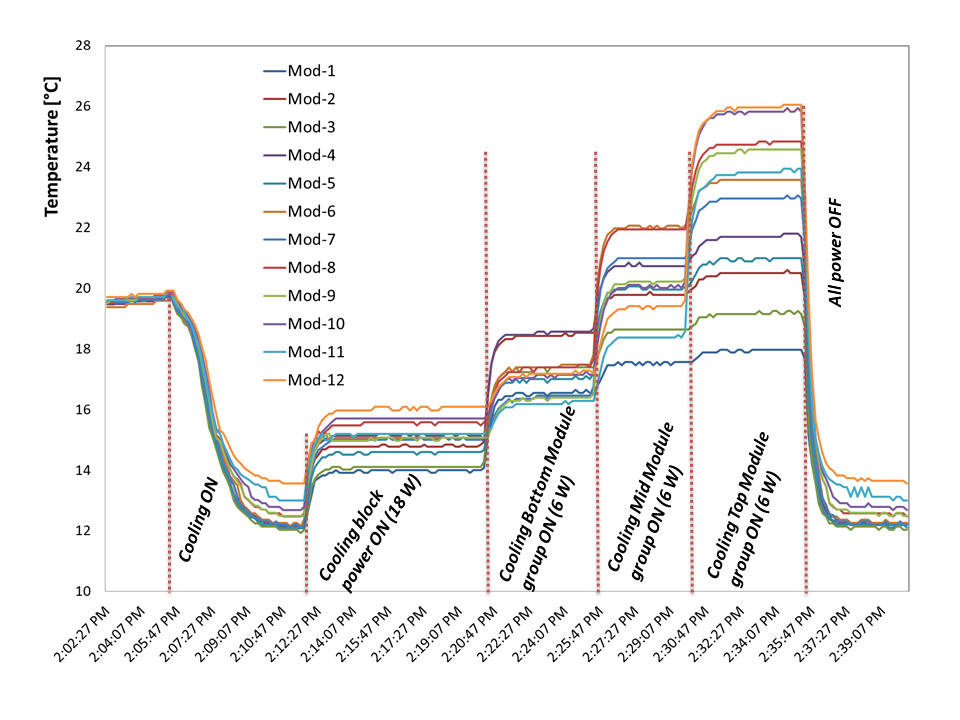}
	\caption{Temperature distribution of the set-up at various steps of the cooling and power configuration (top) as compared to the temperature distribution of the 12 super-modules (bottom).}
	\label{fig:plots}
\end{figure}
The temperature distribution of the super-modules ranges from 18 to \SI{26}{\celsius} at the full power of the heaters (as shown in figure \ref{fig:plots}). The first dummy super-module has the lowest measured temperature because of the effect of the base plate. All the measurements show that the target of a maximum temperature of \SI{40}{\celsius} is achievable for a detector consisting of 16 towers and a power consumption of \SI{300}{\watt}, at the condition that a coolant flow inside the blocks are guaranteed and that the minimum HTC value of \SI{4000}{\watt\per\meter\squared\per\kelvin} is reached.\\
An FEA model, shown in figure \ref{fig:assembly}, was designed to be as close as possible to the real test bench in terms of geometry and materials, including:
\begin{itemize}
	\item 3$\times$4 modules (as described previously)
	\item 1 aluminum cooling block
	\item 1 ceramic \ce{AlO3} cooling block
	\item 1 aluminum base block (assembly jig)
	\item 1 aluminum base plate providing a thermal boundary condition (\SI{19}{\celsius} temperature at the bottom surface)
\end{itemize}
The 12 modules are thermally connected to the side cooling blocks by thermal glue (loaded epoxy or equivalent). The blocks in the mock-up were insulated from the aluminum base block by interfacing Nomex foils. This is accounted for into the FEA model. The simulation was performed using the following setup.\\
Boundary conditions:
\begin{itemize}
	\item The temperature is set at \SI{19}{\celsius} at the bottom side of the base plate in aluminum (constant).
	\item The air convection (\SI{19}{\celsius}) is applied to the side of the cooling blocks (vertical) + the horizontal surfaces of the top module and the aluminum parts (HTC = \SI{1}{\watt\per\meter\squared\per\kelvin})
	\item The coolant (water for inlet/outlet) through the cooling block channels is considered as a convective temperature applied to the surfaces (channels).
\end{itemize}
Loads:
\begin{itemize}
	\item A heat flux is applied to each heater pad, corresponding to \SI{18}{\watt} in total (\SI{2.4}{\milli\watt\per\milli\meter\squared})
	\item 2 heat fluxes are also applied to the cooling blocks sides to account for the effect of adjacent modules (\SI{9}{\watt} per cooling block, equivalent to \SI{0.07}{\watt\per\milli\meter\squared})
\end{itemize}
The coolant temperature at the inlet and outlet of the manifold was set respectively to \SI{11,8}{\celsius} and \SI{16}{\celsius}, in accordance with the measurements on the mock-up. However, since the blocks are quite different (due to heat loss variations and different flows, leading to different HTCs), the temperature of the coolant in each block is expected to be different. The FEA model simulates the two blocks independently in terms of HTC and temperature.
In the first simulation an HTC of \SI{10000}{\watt\per\meter\squared\per\kelvin} for the \ce{AlO3} and \SI{6000}{\watt\per\meter\squared\per\kelvin} for the aluminum block were set. In this condition the maximum temperature difference between the simulation and measurement was of \SI{2}{\celsius}. 
To improve the matching between the simulation and the measurements, a parametric optimization on the two HTC values was performed. The optimal values of the HTCs are respectively \SI{5000}{\watt\per\meter\squared\per\kelvin} for the aluminum and \SI{9000}{\watt\per\meter\squared\per\kelvin} forthe  \ce{AlO3} block, for which the temperature mapping at full power (\SI{36}{\watt}) is shown in figure \ref{fig:temperaturemapping}. The asymmetry between the two cooling blocks is clearly visible: temperatures on the cooling blocks have a discrepancy of 2 to \SI{3}{\celsius} for the same coolant temperature. The results, shown in table \ref{tab:FEA}, show good accordance with the measurements with an average error of less than \SI{1}{\celsius}, offering a good starting point towards simulations of the full scanner.
\begin{table}[htbp]
	\centering
\begin{tabular}{|c|c|c|}
	\hline 
	MOD \# & Measurement Temp (\SI{}{\celsius})& FEA Temp (\SI{}{\celsius})\\ 
	\hline 
	1 & 18.0 & 19.6 \\ 
	\hline 
	2 & 20.5 & 20.6 \\ 
	\hline 
	3 & 19.3 & 22.5 \\ 
	\hline 
	4 & 21.8 & 22.9 \\ 
	\hline 
	5 & 21.0 & 24.0 \\ 
	\hline 
	6 & 23.6 & 24.3 \\ 
	\hline 
	7 & 23.1 & 24.8 \\ 
	\hline 
	8 & 24.8 & 24.8 \\ 
	\hline 
	9 & 24.6 & 25.0 \\ 
	\hline 
	10 & 25.6 & 24.8 \\ 
	\hline 
	11 & 24.0 & 25.0 \\ 
	\hline 
	12 & 26.0 & 24.6 \\ 
	\hline 
\end{tabular} 
\caption{List of measured and simulated temperature on each module of the thermal mock-up}
\label{tab:FEA}
\end{table}
\\
\section{Conclusions}
\label{sec:conclusions}
The TT-PET small animal scanner has an innovative multi-layer layout that presents many challenges in terms of implementation and production. In particular for what concerns the connection of electrical interfaces and services and the cooling of a system of this complexity in which the detector dead area must be minimized. A stacked wire-bond scheme was proposed. Test results on dummy substrates proved successful, showing a reliable and reproducible interconnection technique for the daisy-chained silicon detectors. 3D-printed cooling blocks using different materials were produced and found to be performing according to simulations. Although they are capable of dissipating the power produced by the scanner, a new design is being investigated to improve the yield of the manufacturing process.

\acknowledgments
We would like to thank Prof. Allan Clark for his precious suggestions and for reading this manuscript and the technicians of the DPNC of University of Geneva for their contribution. This study was funded by the SNSF SINERGIA grant CRSII2\_160808. 


\end{document}